\documentclass[11pt,english,aps]{revtex4}
\usepackage[T1]{fontenc}
\usepackage[latin1]{inputenc}
\usepackage{graphicx}
\usepackage{amssymb}

\makeatletter

\providecommand{\tabularnewline}{\\}

\renewcommand{\arraystretch}{1.3}

\usepackage{babel}
\makeatother
\begin{document}

\preprint{September 2006}

\title{Mixed Action Simulations on Staggered Background;\\
Interpretation and Result for the 2-flavor QCD Chiral Condensate}

\author{Anna Hasenfratz}

\email{anna@eotvos.colorado.edu}

\affiliation{Department of Physics, University of Colorado, Boulder, CO-80309-390}

\author{Roland Hoffmann}

\email{hoffmann@pizero.colorado.edu}

\affiliation{Department of Physics, University of Colorado, Boulder, CO-80309-390}

\begin{abstract}
Growing evidence indicates that in the continuum limit the rooted
staggered action is in the correct QCD universality class, the non-local
terms arising from taste breaking can be viewed as lattice artifacts.
In this paper we consider the 2--flavor Asqtad staggered action at
lattice spacing $a\!\approx\!0.13\,$fm and probe the properties of
the staggered configurations by an overlap valence Dirac operator.
By comparing the distribution of the overlap eigenmodes to continuum
QCD predictions we investigate if/when the lattice artifacts are small
as a function of the staggered quark mass. We define a matching overlap
quark mass where the lattice corrections are minimal for the topological
susceptibility and from the eigenmode distribution we predict the
2--flavor chiral condensate. Our results indicate that the staggered
configurations are consistent with 2--flavor continuum QCD up to small
lattice artifacts, and predict a consistent value for the infinite
volume chiral condensate.
\end{abstract}
\maketitle

\section{Introduction}

With the advent of improved actions, increasing dedicated computing
power, and large scale collaborative efforts, lattice QCD calculations
became mainstream, the numerical predictions are often comparable
or better than other theoretical calculations. Most of the recent
high precision results were obtained using staggered fermions. The
2+1 flavor Asqtad configurations created by the MILC/SciDac collaboration
are publicly available and have been used both with staggered valence
quarks and also in mixed action simulations with overlap or domain
wall fermions. 

Staggered fermions are computationally advantageous but the staggered
action describes four fermionic flavors. In order to simulate 2+1
flavor QCD the $4^{\mathrm{th}}$ root of the fermionic determinant
is taken in the Boltzmann weight. This procedure, usually referred
to as rooting, leads to a non-local fermionic action that might not
be in the expected 1--flavor QCD universality class. Non-locality
does not necessarily mean that the action is incorrect. If the non-local
terms are irrelevant, i.e. scale away with the lattice spacing, the
staggered action has the correct continuum limit, but this is not
automatic and cannot be proven by usual renormalizability arguments
\cite{Bunk:2004br,Adams:2004mf,Shamir:2004zc,Hasenfratz:2005ri,Bernard:2006ee,Shamir:2006nj}.
In Ref. \cite{Shamir:2006nj} Shamir has developed a framework, based
on renormalization group considerations, that shows how the non-local
terms of the rooted staggered action can become irrelevant. The proof
relies on several assumptions, and while those assumptions are quite
plausible, their validity has to be checked. 

One possibility to support the arguments of Ref. \cite{Shamir:2006nj}
and the validity of the rooted staggered action is to numerically
verify that the non-local terms become irrelevant in the continuum
limit. In a previous work \cite{Hasenfratz:2006nw,Hoffmann:2006XX}
we carried out such a study for the 2--dimensional Schwinger model.
We compared the determinant of the staggered lattice action to the
determinant of a chiral overlap action with the same flavor number
on staggered configuration ensembles. We showed that the difference
between the lattice formulations, both local lattice artifacts and
non-local terms, scale away with at least $O(a^{2})$ and the continuum
limit can be approached with any physical quark mass, even with massless
quarks. The result implies that the non-local terms of the rooted
action can be considered lattice artifacts when the taste breaking
terms of the staggered action are not too large.

In this paper we develop the techniques necessary to carry out a similar
study in 4--dimensional QCD. Our approach is different from Ref. \cite{Hasenfratz:2006nw}
as the high precision evaluation of the fermionic determinant is not
feasible in 4 dimensions. Rather we develop a more general method
that can be applied to any mixed action simulation.

Mixed action simulations became popular in recent years as they combine
the simulation advantages of a simple sea quark action with the exact
or near exact chiral symmetry of overlap or domain wall valence quarks.
The price to pay, in addition to an internal inconsistency (unitarity
violation), is the complication in the analysis. One option is to
derive and use partially quenched mixed action chiral perturbative
formulae. For staggered sea quarks this approach is particularly cumbersome
(although it can be very effective) as the taste breaking terms of
the staggered action require the introduction of dozens of parameters
in the chiral Lagrangian. It is no longer possible to make physical
predictions from individual configuration ensembles, at the end all
data enters into the chiral fitter that predicts chirally extrapolated
continuum quantities. Another possibility is to match the parameters
of the valence and sea quark actions as well as possible and deal
with any remaining difference as part of the lattice artifacts. Since
the lattice action is characterized by only a few parameters, namely
the lattice spacing and the quark masses, this matching is fairly
simple. If both the sea and valence actions have small lattice artifacts,
their difference after matching should also be small. This latter
approach is useful if the chiral perturbative formulae do not exist
or the numerical data does not allow the fitting of all the parameters,
or if one desires more insight into the physics contained in a particular
set of gauge configurations.

The effectiveness of the latter approach was illustrated in Ref. \cite{Hasenfratz:2006nw}
and while there our main objective was to investigate the validity
of the rooting procedure for staggered quarks, we also showed that,
at least within the 2--dimensional Schwinger model, mixed action simulations
with overlap valance quarks on both 2--flavor (unrooted) and 1--flavor
(rooted) staggered sea quark configurations reproduce the full dynamical
overlap action results if the overlap mass is tuned appropriately.
We illustrated this both for the mass dependence of the topological
susceptibility and the (massive) scalar condensate \cite{Hoffmann:2006XX}.

In this work we report our first results along the same lines in 4--dimensional
2--flavor QCD. Our strategy is simple: we investigate to what extent
configurations generated with rooted staggered fermions describe continuum
QCD. The deviation characterizes both local and non-local lattice
artifacts. If these terms are irrelevant in the continuum limit, they
should scale away as the lattice spacing is decreased at fixed physical
quark mass. We do not yet have numerical data to investigate the continuum
limit, in this work we concentrate on the mass dependence at fixed
lattice spacing. We show that configurations generated with 2--flavor
staggered quarks at a single lattice spacing but at four different
quark masses are consistent with 2--flavor QCD configurations, at
least for the three heavier masses. We also show how the topological
charge distributions can be used to determine the best overlap valence
matching masses and that with these mass values the data sets predict
a consistent value for the chiral condensate of 2--flavor QCD. 

In Sect. \ref{sec:Strategy-and-Simulation} we describe our strategy
and summarize our numerical setup. The fit of the eigenvalue distributions
of the Dirac operator to random matrix theory predictions is discussed
in Sect. \ref{sec:Eigenvalues-of-the}, and Sect. \ref{sec:Topology}
describes the topological charge distribution. Finally, in Sect. \ref{sec:The-chiral-condensate}
we combine the previous results to obtain a prediction for the chiral
condensate. In this section we also discuss the dependence of the
matching valence quark mass on the staggered sea mass.

\section{Strategy and Simulation setup\label{sec:Strategy-and-Simulation}}

We generated configuration sets with the rooted staggered action at
fixed lattice spacing and at several quark mass values and ask two
very basic questions:

\begin{enumerate}
\item Are the staggered configuration sets consistent with $n_{f}\!=\!2$
continuum QCD, up to small lattice artifacts?
\item What is the matching valence overlap quark mass that best describes
the staggered configurations?
\end{enumerate}
To investigate the first question one has to consider observables
that are sensitive to the vacuum and do not depend strongly on the
valence quark mass. Spectral quantities are not appropriate, but the
low lying infrared eigenmodes of the massless valence Dirac operator
are a good choice. Another quantity, which we will consider, is the
topological charge of the configurations.

The answer to the second question is not unique. In fact any {}``reasonable''
choice (i.e. one that approaches the continuum fixed point similarly
to the staggered sea quark mass) will work in the continuum limit.
However at finite lattice spacing a wise choice of the matching valence
mass could significantly reduce lattice artifacts. As it turns out
the eigenmodes of the massless Dirac operator are not very sensitive to the
sea quark mass but we will be able to use the topological charge distribution
to fix the matching valence quark mass.

In our calculation we use an overlap valence quark action, based on
an improved Wilson action kernel (planar action) with HYP smeared
links \cite{Hasenfratz:2001hp}. This is the same action that was
used by DeGrand in previous studies \cite{DeGrand:2000tf,DeGrand:2003in}.

Our sea quark action is the 2--flavor Asqtad staggered action \cite{Orginos:1999cr,Bernard:2001av,Aubin:2004wf}.
We have generated four configuration sets, each consisting of 400-500
$12^{4}$ lattices at a lattice spacing of about $a=0.13\,$fm. We
have used the Refreshed Molecular Dynamics algorithm from the publicly
available MILC code %
\footnote{http://www.physics.utah.edu/\textasciitilde{}detar/milc/%
} and chose the step size according to Ref. \cite{Bernard:2001av}.
The details of the sets are summarized in Table \ref{cap:Parameters-of-the}.
The $\beta=7.2$, $am_{st}=0.02$ (\textbf{M}) set corresponds to
the MILC single 2--flavor run.

\begin{figure}[b]
\begin{centering}\includegraphics[bb=136bp 298bp 453bp 536bp,clip,width=75mm]{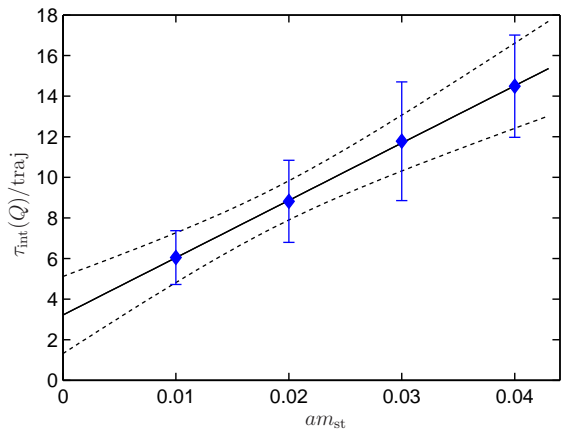}\par\end{centering}

\caption{Integrated autocorrelation time of the topological charge vs. staggered
quark mass. \label{cap:Autocorrelation}}
\end{figure}

The lattice spacing is obtained from the hadronic length scale $r_{0}$
\cite{Sommer:1993ce,Hasenfratz:2001tw}, which we measured on $12^{3}\times16$
configurations. The values of $r_{0}/a$ are listed in Table \ref{cap:Parameters-of-the}
together with the simulation parameters. We also quote the level of
taste breaking, the ratios of the heaviest and lightest pion masses,
approximated from corresponding 2+1 flavor results \cite{Bernard:2001av}.
The last column lists the separation of the configurations in terms
of unit length molecular dynamics trajectories. These numbers were
chosen to reflect the autocorrelation time of the topological charge.

We were surprised to find that the heavier mass data sets had larger
autocorrelation times as shown in Fig. \ref{cap:Autocorrelation}.
We will discuss this problem further at the end of Sect. \ref{sec:Topology}.\def\arraystretch{1.3}

\begin{table}
\begin{tabular}{|c|c|c|c|c|c|}
\hline 
$\ $Set$ $&
$\quad\beta\quad$&
$\ am_{st}\ $&
$\ \ \ r_{0}/a\ \ \ \ $&
$\ $Taste breaking$\ $&
$\ $Time separation$\ $\tabularnewline
\hline
\hline 
\textbf{L}&
7.18&
0.01&
3.84(6)&
60\%&
5\tabularnewline
\hline
\textbf{M}&
7.20&
0.02&
3.82(3)&
34\%&
5\tabularnewline
\hline 
\textbf{H}&
7.22&
0.03&
3.60(4)&
24\%&
10\tabularnewline
\hline
\textbf{E}&
7.24&
0.04&
3.64(3)&
18\%&
15\tabularnewline
\hline
\end{tabular}

\caption{Parameters of the $n_{f}\!=\!2$ staggered background configurations.
The taste breaking levels are approximated from corresponding 2+1
flavor runs except for the \textbf{M} set where 2--flavor spectroscopy
data exists. The molecular dynamics time separations between the configurations
reflect the autocorrelation of the topological charge. \label{cap:Parameters-of-the}}
\end{table}

\section{Eigenvalues of the Dirac Operator and Random Matrix Theory\label{sec:Eigenvalues-of-the} }

\def\arraystretch{1.0}Random matrix theory (RMT) captures the universal
chiral properties of QCD and predicts the distribution of the physical
(infrared) eigenvalues of the massless Dirac operator in the $\epsilon$--regime
\cite{Verbaarschot:2000dy}. The predictions are given in fixed topological
sector $\nu$ and depend on the low energy constant $\Sigma$, the
infinite volume chiral condensate. The distribution of the (microscopically
rescaled) $n^{\mathrm{th}}$ eigenmode $\lambda\Sigma V$ is given
as \begin{equation}
P_{\nu,n}(\lambda\Sigma V)=\Lambda_{\nu,n}(m\Sigma V;n_{f})\;,\label{RMT-pred}\end{equation}
where $m$ is the sea quark mass and $V$ is the volume of the configurations.
$\Lambda_{\nu,n}$ is an $n_{f}$ dependent universal function of
the variable $M=m\Sigma V$. While $\Lambda_{\nu,n}$ can in principle
be obtained directly from random matrices, analytical forms also exist
and are more convenient for evaluating the distribution of the lower
modes \cite{Damgaard:2000ah,Damgaard:2000qt}. While the universal
predictions of RMT hold strictly only in the $\epsilon$--regime and
in infinite volume, in practice the validity of random matrix theory
seems to extend significantly further \cite{Giusti:2003gf,DeGrand:2006qu}.

Since the eigenvalues $\lambda_{\nu,n}$ refer to the massless Dirac
operator, the quark mass dependence enters only through the sea quarks.
In dynamical simulations, where the same chiral action is used both
for the sea and valence sectors, $m$ in Eq.(\ref{RMT-pred}) is known
and the eigenvalue distribution can be used to predict the chiral
condensate $\Sigma$ \cite{DeGrand:2006uy,DeGrand:2006nv}. In our
case $m$ is an overlap quark mass that corresponds to the background
configurations that were generated by staggered quarks, i.e. $m$
is the matching quark mass as described in Sect. \ref{sec:Strategy-and-Simulation}.
The value of $m$ is not known a priori and therefore in our case
$P_{\nu,n}(\lambda)$ depends on two variables, $M$ and $\Sigma$.
We fit the measured eigenvalue distribution to random matrix theory
at fixed $M$ and predict the chiral condensate $\Sigma(M)$. The
systematic deviation of the data from the RMT prediction of Eq.(\ref{RMT-pred})
characterizes the lattice artifacts, both from discretization errors
and from the non-locality of the action. This deviation is the measure
of consistency between the lattice action and continuum QCD and replaces
the residue used in Ref. \cite{Hasenfratz:2006nw} for the same purpose.
If the rooting procedure is correct, it should scale to zero as the
continuum limit is approached at fixed physical (matching) quark mass,
assuming the simulations are done in the region where the RMT predictions
are valid.

It is customary to fit the integrated or cumulative eigenvalue distribution,
which avoids the binning of the data. The most frequently used fit
is the Kolmogorov-Smirnov (KS) test that minimizes $D_{\mathrm{max}}^{2}$,
the maximal deviation between the measured and the predicted cumulative
distributions \cite{Bietenholz:2003mi,DeGrand:2006uy,DeGrand:2006nv}.

An advantage of the KS test is that there is an explicit and simple
form for the confidence level of the fit that gives the probability
that the measured distribution is consistent with the analytical one.
This so called quality factor \begin{eqnarray}
Q_{\mathrm{KS}}(d) & = & 2\sum_{j=1}^{\infty}(-1)^{j\!+\!1}e^{-2j^{2}d^{2}},\nonumber \\
d & = & (\sqrt{N}+0.12+0.11/\sqrt{N})D_{\mathrm{max}}\label{eq:Q_factor}\end{eqnarray}
depends on $D_{\mathrm{max}}$ and $N$, the number of configurations
in the sample. For our data sets the quality factor is well approximated
by the first term in the sum, \begin{equation}
\log Q_{\mathrm{KS}}\approx\log2-2ND_{\mathrm{max}}^{2}\;.\label{eq:}\end{equation}
The KS fit maximizes the quality factor $Q_{\mathrm{KS}}$ or the
product of quality factors if more than one distribution is used.
However, $Q_{\mathrm{KS}}$ will go to zero exponentially with increasing
statistics if the measured distribution is not \emph{exactly} described
by the analytic form. In any lattice calculation there are lattice
artifacts and finite volume effects, so the analytic form is never
exactly reproduced, the quality factor vanishes as the numerical statistics
increases. Quality factors quoted in this situation are rather meaningless,
unless one uses it to compare simulations with identical statistics.
On the other hand $D_{\mathrm{max}}$ has a finite limit as the statistics
increases and describes the systematic deviation of the numerical
and analytical values. In the following we fit our data by maximizing
the quality factor (or products of quality factors) according to the
KS test but describe the goodness of the fit by the value $D_{\mathrm{max}}$
itself.

\begin{figure}
\includegraphics[scale=0.95]{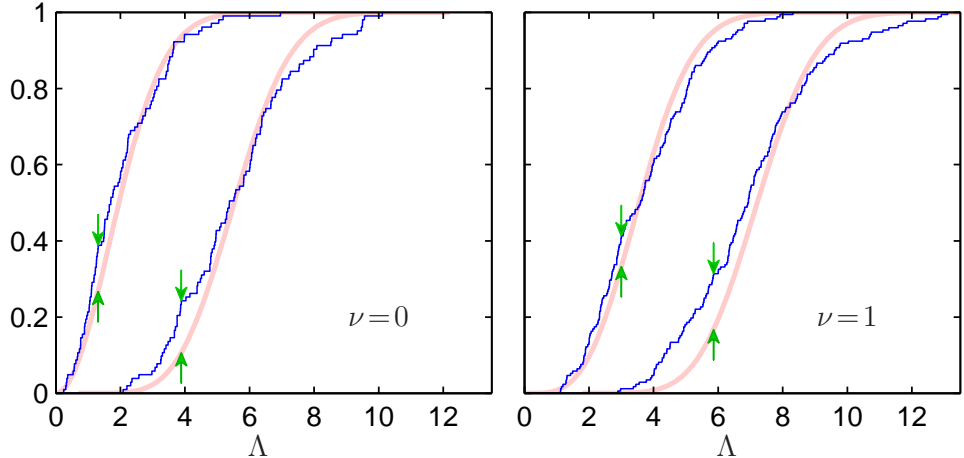}

\caption{RMT predictions of the cumulative distribution of the two lowest
eigenmodes in the $\nu=0$ and $1$ sectors of the \textbf{M} ($am_{st}\!=\!0.02)$
data set at $M\!=\! m\Sigma V\!=\!13.5$ (see below). The fit uses
only the first modes of the $\nu=0$ and 1 topological sectors. Arrows
indicate the maximal deviation between the data set and the analytical
predictions. \label{cap:RMT-fit}}
\end{figure}

The random matrix theory predicts the distribution of the universal,
infrared eigenmodes. A rough estimate for the number of such modes
comes from the expected number of instantons on each configurations.
Since our volume is about 6$\,$fm$^{4},$ we expect on average 6
instantons per configurations. That implies that only the first 1-2
modes in the $\nu=0-2$ sectors can be expected to be infrared dominated.
Somewhat arbitrarily we have decided to fit the first modes in the
$\nu=0$ and 1 sectors. Figure \ref{cap:RMT-fit} shows some typical
fits of the cumulative distribution for the \textbf{M} ($am_{st}=0.02)$
data set. The left panel corresponds to the $\nu=0$, the right panel
to the $\nu=1$ sector. The first modes of the $\nu=0$ and 1 topological
sectors are included in the fit. In addition to the two fitted modes
we also show the non-fitted second modes in the same topological sectors.
$D_{\mathrm{max}}$ is almost a factor of two smaller for the $n=1$,
$\nu=1$ mode, but not significantly worse for the non-fitted modes
than for the fitted $n=1$, $\nu=0$ mode.

\begin{figure}
\includegraphics[scale=0.8]{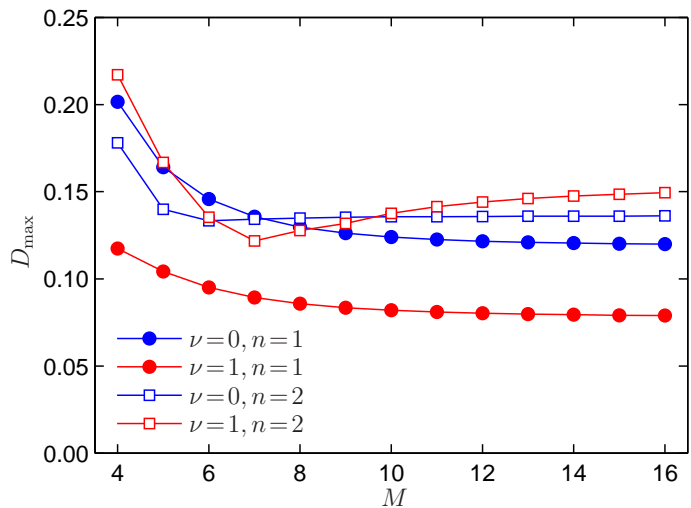}

\caption{The maximal deviation $D_{\mathrm{max}}$ as the function of $M=m\Sigma V$
for the \textbf{M} data set. The fit is to the first modes of the
$\nu=0$ and 1 sectors (filled points). The second modes in the same
topological sectors are not fitted (open symbols). \label{cap:D_max vs M}}
\end{figure}

In Fig.\ref{cap:D_max vs M} we plot the maximal deviations $D_{\mathrm{max}}$
as a function of the RMT parameter $M$. Evidently the quality of
the fit is not very sensitive to this parameter. While small values
are disfavored -- $D_{\mathrm{max}}$ increases by a factor of 2 if
$M\lesssim4$ -- larger values are almost equally probable. Contrary
to our original hopes the eigenmode distributions cannot be used to
define a matching mass, it defines only a range of acceptable values.
It is interesting to note that the same analysis on the dynamical
overlap data set of Ref. \cite{DeGrand:2006nv} shows a broad but
definite minimum in $D_{\mathrm{max}}$ around the actual sea quark
mass %
\footnote{We thank the authors of Ref. \cite{DeGrand:2006nv} for sharing their
raw data with us.%
}. 

\begin{figure}
\includegraphics{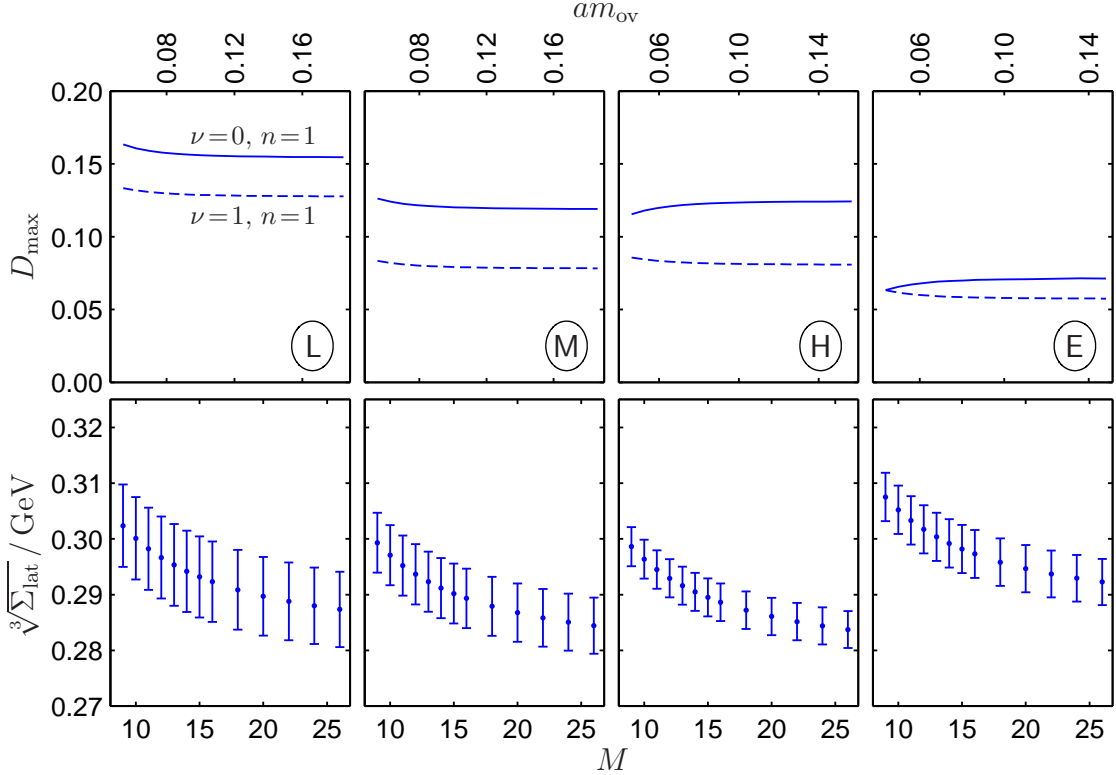}

\caption{$D_{\mathrm{max}}$ and $\Sigma^{1/3}$ in GeV as the function of
$M=m\Sigma V$ for all four data sets. \label{cap:Sigma for all}}
\end{figure}

The result of the fit is similar for the other three data sets. The
upper panels of Figure \ref{cap:Sigma for all} show $D_{\mathrm{max}}$
for the fitted modes, and the dependence on the staggered mass is
obvious. $D_{\mathrm{max}}$ is significantly lower at the heaviest
\textbf{E} data set than for the lightest \textbf{L} one, with the
intermediate mass sets lying in between. This behavior is expected
since at finite lattice spacing a smaller staggered mass leads to
increased taste symmetry breaking (Table \ref{cap:Parameters-of-the}),
it differs more from the flavor symmetric valence quark sector. With
decreasing lattice spacing at fixed physical quark mass this deviation
should decrease and eventually vanish in the continuum limit.

At each $M=m\Sigma V$ value the fit predicts $\Sigma V/a$. Using
the values for $r_{0}/a$ from Table \ref{cap:Parameters-of-the}
this can be converted to physical units as shown on the lower panels
of Fig. \ref{cap:Sigma for all}. The corresponding $am=M/(\Sigma V/a)$
overlap mass values are shown along the upper border of the figure.
For those configuration sets that are consistent with continuum QCD
the predicted $\Sigma$ values should be identical, independent of
the sea quark mass. Unfortunately at this point we do not know the
matching $M$ values to predict $\Sigma$. However we can already
exclude a simple linear $m=Zm_{st}$ relation with $Z=O(1)$. Such
a relation would require that the \textbf{L} and \textbf{E} data sets
predict the same condensate at $M$ values that are a factor of 4
different. That is obviously not consistent with the data unless $Z$
is unreasonably large. In order to predict the chiral condensate we
have to find an independent quantity that can be used to fix the matching
valence quark mass. The topological charge distribution is a possible
choice as we will discuss in Sect. \ref{sec:Topology}.

\def\arraystretch{1.2}

\begin{table}
\begin{tabular}{|c|c|c|c|c|r|c|c|}
\hline 
$\ $Set$\ $&
$\ M=m\Sigma V\ $&
$\ \Sigma^{1/3}$ / MeV$\ $&
$am$&
$\ \nu\ $&
$\ N\ $&
$\ D_{{\rm max}}\ $&
$\ Q_{{\rm KS}}\ $\tabularnewline
\hline
\hline 
\textbf{L}&
12.7(2.0)&
295.7(7.0)&
0.083(4)(14)&
0&
89&
0.157&
0.022\tabularnewline
\hline 
\multicolumn{4}{|c|}{}&
1&
144&
0.130&
0.014\tabularnewline
\hline 
\textbf{M}&
13.5(2.8)&
291.7(4.1)&
0.090(3)(19)&
0&
103&
0.121&
0.092\tabularnewline
\hline 
\multicolumn{4}{|c|}{}&
1&
172&
0.080&
0.217\tabularnewline
\hline 
\textbf{H}&
16.9(1.9)&
288.0(5.4)&
0.098(4)(12)&
0&
87&
0.123&
0.132\tabularnewline
\hline 
\multicolumn{4}{|c|}{}&
1&
178&
0.081&
0.181\tabularnewline
\hline 
\textbf{E}&
22.6(4.3)&
293.5(4.1)&
0.127(4)(24)&
0&
85&
0.071&
0.768\tabularnewline
\hline 
\multicolumn{4}{|c|}{}&
1&
193&
0.058&
0.022\tabularnewline
\hline
\end{tabular}

\caption{Results of the RMT fit to the lowest eigenmodes in the $\nu=0,1$
sectors. $N$ is the number of configurations of each data set, $D_{\mathrm{max}}$
is the maximum deviation of the fit and the quality factor $Q_{\mathrm{KS}}$
is defined in Eq. (\ref{eq:Q_factor}). $am$ is the matching overlap
mass where the first error is due to the uncertainty of $\Sigma$
only, while the second one takes into account both the errors of $\Sigma$
and $M$. For the determination of $M$ see Sect. \ref{sec:Topology}.\label{cap:Results}}
\end{table}

\def\arraystretch{1.0}In Table \ref{cap:Results} we list the number
of configurations we had in the $\nu=0$ and 1 topological sectors,
the $D_{\mathrm{max}}$ values of the RMT fit at specific $M=m\Sigma V$
values and the corresponding quality factors. The choice of these
$M$ parameters will be explained in Sect. \ref{sec:Topology}. Are
these fits and corresponding $D_{\mathrm{max}}$ values reasonable?
All our data have been obtained at a single lattice spacing so we
cannot investigate the lattice spacing dependence. In order to develop
a feel for the quality of the fit we looked at published data both
for quenched and dynamical simulations. In Ref. \cite{Bietenholz:2003mi}
the $D_{\mathrm{max}}$ values for the first mode on physically slightly
smaller $12^{4}$ lattices are 0.27 and 0.08 for $\nu\!=\!0$ and
1, respectively. This study used an improved but not smeared overlap
operators on quenched lattices with somewhat smaller lattice spacing
than ours. In Ref. \cite{DeGrand:2006nv} the $D_{\mathrm{max}}$
values of the first non--zero modes using the same fitting strategy
as ours are between 0.11 and 0.20. That study used dynamical overlap
configurations with the same improved overlap operator we use here
with two levels of stout smeared links at a somewhat coarser lattice
spacing. In view of these numbers we can conclude that, as far as
the Dirac operator eigenmode distribution is concerned, the rooted
staggered action configurations do not show larger lattice artifacts
than the overlap ones. Even the worst set with the lightest quark
mass is comparable to the overlap data.

\section{Topology\label{sec:Topology}}

We have seen in Sect. \ref{sec:Eigenvalues-of-the} that the low energy
eigenmodes of the massless overlap Dirac operator are consistent with
continuum QCD as predicted by RMT as long as the quantity $M=m\Sigma V$
is larger than some minimal value. The RMT fit however does not restrict
$m$ any further. A matching along the lines of Ref. \cite{Hasenfratz:2006nw},
with the determinants approximated by a small number of infrared modes,
can in principle be used to fix the valence quark mass. While this
method gives masses in the expected range, the result varies with
the number of included modes (we measured 8 overlap and 32 staggered
eigenvalues). Unless more eigenmodes are available, this leads to
unacceptably large errors in the overlap mass. We therefore need to
consider another physical observable in order to match $m$ to the
dynamical configurations. Again, we want to use a quantity that is
sensitive to the vacuum structure and not the valence quark sector,
so we chose the topological charge.

Since we have sufficient statistics, over 400 approximately independent
configurations at each coupling value on not too large volumes ($12^{4}$
or about 6 fm$^{4}$), we can study the topological charge distribution.
Following the discussion of Refs. \cite{Leutwyler:1992yt,Verbaarschot:2000dy,Durr:2001ty},
we write the probability of encountering a charge $\pm\nu$ configuration
in the dynamical ensemble as \begin{equation}
P_{\nu}=Z_{\nu}(m\Sigma V)Q_{\nu}(\sigma)\;.\label{eq:Top-prob}\end{equation}
Here $Q_{\nu}$ is the quenched probability of a charge $\pm\nu$
configuration while $Z_{\nu}$ describes the suppression due to the
fermionic determinant. Based on simple probabilistic arguments $Q_{\nu}$
is expected to be a Gaussian distribution. Recent large scale simulations
support this expectation in large volumes \cite{Giusti:2003gf}. The
data is consistent with \begin{equation}
Q_{\nu}(\sigma)=\frac{1}{\sqrt{2\pi\sigma^{2}}}e^{-\nu^{2}/2\sigma^{2}}\{1+O(V^{-1})\}\;,\label{eq:Top-Quen}\end{equation}
where $\sigma^{2}=\langle\nu^{2}\rangle_{Q}=V\chi_{Q}$ is the expectation
value of the charge squared in the quenched theory. The fermionic
suppression factor has been calculated both within chiral perturbation
theory and the random matrix model \cite{Leutwyler:1992yt,Verbaarschot:2000dy}.
For 2 flavors it is \begin{equation}
Z_{\nu}=I_{\nu}^{2}(M)-I_{\nu-1}(M)I_{\nu+1}(M)\;,\label{eq:Top-Dyn}\end{equation}
where the $I_{\nu}(M)$ are modified Bessel functions. Thus the charge
probability distribution $P_{\nu}$ depends on two variables, $M=m\Sigma V$
and $\sigma$. The latter can be determined from the quenched topological
susceptibility, so a one parameter fit to the topological charge distribution
data predicts $M$.

\begin{figure}
\includegraphics[scale=0.75]{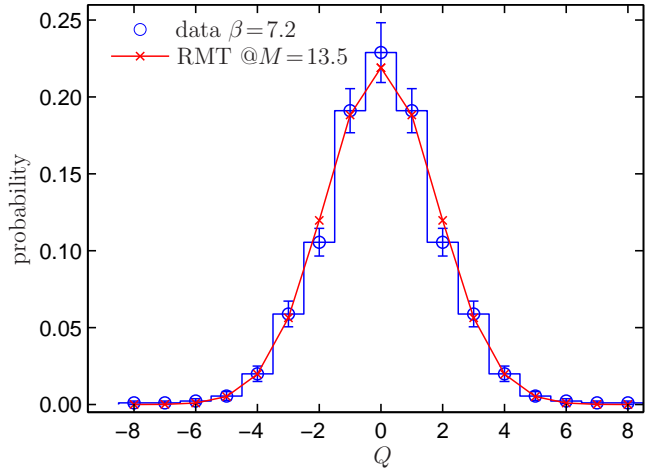}

\caption{The topological charge distribution for the \textbf{M} data set with
fit to Eq.(\ref{eq:Top-prob})\label{cap:The-topological-charge}.}
\end{figure}

\begin{figure}
\includegraphics[scale=0.85]{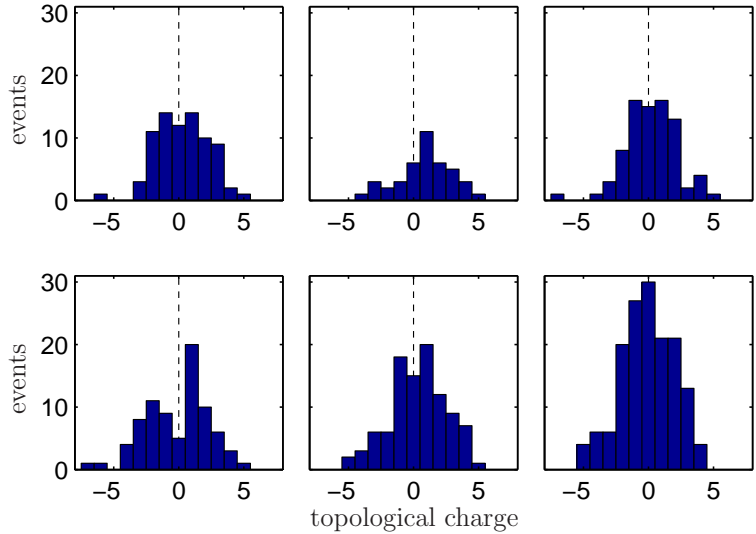}

\caption{Histograms of the topological charge for the \textbf{E} data set,
split up by independent series.\label{cap:strange-effect}}
\end{figure}

Figure \ref{cap:The-topological-charge} shows such a fit for the
\textbf{M} configuration set using $\chi_{Q}r_{0}^{4}=0.072$ from
Ref. \cite{Giusti:2003gf}. The fit predicts $M=13.5(2.8)$ with $\chi^{2}/\mathrm{DoF}=0.65$.
The predicted $M$ values for the other data sets are listed in Table
\ref{cap:Results}. While the quality of the fit is very similar for
the lightest \textbf{L} data set, for heavier masses we encountered
difficulties in sampling the topological charge distribution due to
long autocorrelation times. Even with increased separation between
consecutive configurations the charge distribution does not look Gaussian.
Both in the \textbf{H} and \textbf{E} data set the $\nu=0$ sector
is under-represented. This probably implies that we still underestimate
the autocorrelation time at the larger quark masses. While the topological
charge value changes from configuration to configuration and covers
both negative and positive values, we have observed systematic shifts
from zero average that persist over tens of configurations. This possibly
could signal an occasional large topological object that is destroyed
only very slowly, but is accompanied by many frequently changing smaller
objects. We are not sure if this is the consequence of the inexact
R algorithm or similar problems would be encountered in exact molecular
dynamics simulations as well. As an illustration Fig. \ref{cap:strange-effect}
shows the distribution of the topological charge on the \textbf{E}
set that contains 550 configurations. In this set we had 6 independent
series with between 40 and 150 consecutive configurations each. The
configurations were separated by 15 molecular dynamics trajectories.
While some sets show a reasonable distributions, others are obviously
biased in one or the other direction. The issue of the autocorrelation
time certainly deserves further consideration.

\section{The chiral condensate and matching masses\label{sec:The-chiral-condensate}}

With the $M=m\Sigma V$ values predicted from the topological charge
distribution we are now able to extract the physical value of the
chiral condensate. Combining the $M$ values with $r_{0}/a$ from
Table \ref{cap:Parameters-of-the} we find that all four configuration
sets predict a consistent value for the $\Sigma$ condensate, as listed
in Table \ref{cap:Results}. The only sign that the light \textbf{L}
set differs from the RMT prediction more than the other mass values
is the larger error of the predicted condensate. The value we obtain,
\begin{equation}
\Sigma_{\mathrm{lat}}^{1/3}=291(5)\,\mathrm{MeV}\label{eq:Cond}\end{equation}
 is the lattice condensate. To connect it to a more conventional scheme,
like $\overline{\mathrm{MS}}$ at 2$\,$GeV, one needs the corresponding
renormalization factor $Z_{s}$. Such a factor should be calculated
non-perturbatively on the staggered configurations with our specific
valence Dirac operator. We have not done this calculation yet but
similar ones exist. The same Dirac operator was used on a quenched
data set in \cite{DeGrand:2005af}, while a similar one with stout
instead of HYP smearing on $n_{f}=1$ and 2 dynamical configurations
was used in \cite{DeGrand:2006uy,DeGrand:2006nv}. $Z_{s}$ seems
to be largely independent of the detailed properties of the background
configurations and we estimate its value to be \begin{equation}
Z_{s}\geq0.9\;,\label{eq:Z_s}\end{equation}
which could lower the value of $\Sigma^{1/3}$ by 3\%. In addition,
there is a finite volume correction that could be rather large, up
to 10-15\% for $\Sigma^{1/3}$ \cite{Damgaard:2001ep,DeGrand:2006qu},
further decreasing the value of the condensate. These effects will
have to be investigated but they are beyond the scope of the present
paper. 

The value of the condensate in Eq.(\ref{eq:Cond}) is consistent with
predictions obtained on overlap dynamical configurations \cite{DeGrand:2006nv}.
The agreement further supports our observation that the rooted staggered
configurations are QCD like, the non-local terms of the action can
be simply taken into account as lattice artifacts. Of course to really
support this statement one has to repeat the calculation at different
lattice spacings. Such a study is under way and its result will be
reported separately.

Finally we turn our attention to the matching overlap mass values.
Combining $M$ and $\Sigma$ we get the values listed in the ''$am$''
column of Table \ref{cap:Results}. These matching masses are not
only surprisingly large but they do not depend linearly on the staggered
masses. While the staggered quark mass changes a factor of four between
the lightest and heaviest data sets, the matching overlap masses change
only 50\% . This is similar to what we observed in the Schwinger model
\cite{Hasenfratz:2006nw}, the matching valence masses show an overall
shift compared to the staggered sea mass values. In addition at very
small sea quark masses, where the matching breaks down, the valence
quark masses are largely independent of the sea quark mass values.
This is illustrated in Fig. 3 of Ref. \cite{Hasenfratz:2006nw}. Such
behavior implies that staggered configurations at small quark masses
are not necessarily closer to chiral continuum QCD than the heavier
mass configurations. All the computational efforts creating light
configurations could be in vain, creating only configurations with
larger lattice artifacts. This might not be a problem when the data
is analyzed with the whole machinery of staggered partially quenched
chiral perturbation theory but should be considered when individual
configuration sets are analyzed in mixed action simulations. Of course
this is only a lattice artifact and any such effect will disappear
as the continuum limit is approached. 

As a final comment we note that staggered chiral perturbation theory
indicates that at leading order the topological susceptibility depends
on the taste singlet pion \cite{Billeter:2004wx}. That is the heaviest
of the pseudoscalars and it is quite conceivable that our matching
procedure identified a quark mass corresponding to this pion. To test
that assertion one would have to measure the overlap pion spectrum
on the staggered background configurations.

\section{Conclusion}

We have studied the properties of the rooted staggered action in a
mixed action simulation using overlap valence quarks. By comparing
physical quantities that are independent of the valence quark mass
to continuum QCD predictions we can identify lattice artifacts and
study their dependence on the lattice spacing and sea quark masses.
In this work we considered the eigenvalue distribution of the massless
Dirac operator and the distribution of the topological charge. We
compared the former to the universal predictions of random matrix
theory and found that the systematic deviation of the data from the
predictions were comparable to quenched and dynamical overlap simulations,
especially at larger sea quark masses. Using the topological charge
distribution we could identify the matching overlap valence quark
mass value which best describes the staggered configurations. We found
these matching values to be fairly large and their dependence on the
staggered sea mass values is not consistent with a simple linear renormalization
factor. With the use of the matching mass we extracted the value of
the chiral scalar condensate. We found that the predictions from all
of our staggered configuration sets were consistent. These findings
indicate that at our lattice spacing, $a\!\approx\!0.13\,$fm, and
with not very light sea quarks the rooted staggered lattice configurations
have lattice artifacts similar to other lattice action, the non-local
terms arising from the rooting procedure can be simply considered
as part of the cutoff effects. In order to show that these non-local
terms indeed become irrelevant in the continuum limit the calculation
have to be repeated at different lattice spacings and the scaling
of the lattice artifacts investigated. It would also be important
to study in a similar manner the lattice artifacts of other observables.

\begin{acknowledgments}
We thank T. DeGrand for the use of his overlap and eigenvalue codes.
Discussions with P. Damgaard, T. DeGrand, S. Schaefer and P. Weisz
are gratefully acknowledged. This research was partially supported
by the US Dept. of Energy.\bibliographystyle{apsrev}
\bibliography{lattice}

\end{acknowledgments}

\end{document}